\documentstyle[epsfig]{mn}

\begin{document}

\title[Black hole mass -- galaxy age relation]{The black hole mass -- galaxy age
relation}

\author[Merrifield, Forbes \& Terlevich]{
 M.R. Merrifield$^1$, 
 Duncan A. Forbes$^{2,3}$ and A.I.~Terlevich$^2$\\
 $^1$School of Physics and Astronomy, University of Nottingham, Nottingham NG7 2RD\\
 $^2$School of Physics and Astronomy, University of Birmingham, Birmingham B15 2TT\\
 $^3$Astrophysics \& Supercomputing, Swinburne University, Hawthorn VIC 3122, 
 Australia}

\date{Received:\ \ \ Accepted: }
 
\pagerange{\pageref{firstpage}--\pageref{lastpage}}%

\maketitle

\label{firstpage}

\begin{abstract}
We present evidence that there is a significant correlation between
the fraction of a galaxy's mass that lies in its central black hole
and the age of the galaxy's stellar population.  Since the absorption
line indices that are used to estimate the age are luminosity
weighted, they essentially measure the time since the last significant
episode of star formation in the galaxy.  The existence of this
correlation is consistent with several theories of galaxy formation,
including the currently-favoured hierarchical picture of galaxy
evolution, which predicts just such a relation between black hole mass
and the time since the last burst of merger-induced star formation.
It is not consistent with models in which the massive black hole is
primordial, and hence uncoupled from the stellar properties of the
galaxy.

\end{abstract}


\section{Introduction}

The existence of active galactic nuclei has long been taken as
evidence for the existence of massive black holes in the centres of
some galaxies (Lynden-Bell 1969).  However, it is only relatively
recently that high spatial resolution studies of the kinematics of
galactic nuclei have revealed that essentially all galaxies harbour
large central masses [see Ho (1999) for a review of the evidence].
The existence of these observations also means that there are now
enough data to study the demographics of massive black holes, in order
to seek clues to their origins.

The first significant discovery in this regard is that there is a
correlation between the mass of the black hole, $M_{\rm BH}$, and the
mass of the host galaxy's spheroidal component,
\footnote{The term
``spheroidal component'' refers to the whole system in the case of
elliptical galaxies, but just the bulge in systems with significant
disk components.}  
$M_{\rm sph}$.  Although there is a variety of
possible biasses in measuring this correlation, it seems broadly to be
the case that there is a linear relationship, such that $M_{\rm BH}
\sim 0.005 M_{\rm sph}$ (Magorrian et al.\ 1998).

Although this correlation is reasonably strong, there is still
considerable scatter in the relation, such that there is more than a
factor of ten variation in the inferred value of $M_{\rm BH}$ for
galaxies of given spheroid mass (Magorrian et al.\ 1998).  Some of
this scatter can probably be attributed to the uncertainties in
calculating black hole masses from relatively poor kinematic data and
simplified dynamical models (van der Marel 1997).  However, there are
also astrophysical reasons why one might expect significant dispersion
in this relation.  For example, consider the simplest possible scenario
in which galaxies form and evolve in near isolation.  If the central
black holes in these galaxies accrete mass fairly steadily from their
hosts, then the mass of a black hole simply reflects the age of its
host.

Under the currently-favoured hierarchical paradigm for galaxy
formation, in which larger galaxies are formed from the merging of
smaller galaxies (White \& Rees 1978), the simple linear correlation
between galaxy mass and black hole mass is readily explained.  Each
time two galaxies merge to form a larger system, their black holes
rapidly spiral to the centre of the new galaxy due to dynamical
friction.  The black holes then merge, creating a
proportionately-larger black hole.  However, a galaxy formed by this
process of repeated mergers cannot be characterized by a single age,
so the above explanation for the scatter in black hole masses must be
modified somewhat.  One measure of such a galaxy's age is the time
since it last underwent a major merger, and Kauffmann \& Haehnelt
(2000) have shown that this timescale is a key factor in explaining
the scatter in black hole masses.  If the last merger happened long
ago, then it will have occurred between relatively unevolved galaxies
in which there would have been a large amount of cold gas.  If the
black hole accretes some fixed fraction of this gas, then galaxies in
which the last merger occurred longer ago will contain more massive
black holes.

This picture, in which black holes acquire much of their mass through
accretion of material from their host galaxies, seems quite credible.
However, it is not the only possible scenario.  Stiavelli (1998) has
argued that galaxies with essentially identical properties could be
formed around pre-existing massive black holes.  In this case, the
spread in black hole masses would simply reflect the stochastic nature
of whatever physics was responsible for the formation of the
primordial black holes.

In this Letter, we investigate whether we can attribute the observed
scatter in black hole masses to an astrophysical cause, and hence
whether we can discriminate between the above scenarios.
Specifically, we investigate whether the masses of black holes
correlate with the ages of their host galaxies as determined by
stellar absorption line diagnostics.  
%Since these age indicators
%measure a luminosity-weighted mean age, they provide an estimate of
%the time since the last significant burst of star formation.  Thus,
%the presence of a correlation between this measure of age and black
%hole mass would provide direct evidence for a scenario in which black
%hole growth is coupled to the major galaxy mergers that induce bursts
%of star formation.

\section{Analysis}

\subsection{Black hole mass determinations}

There are now several compilations of central black hole mass
estimates in nearby galaxies (e.g.\ Ho 1999).  The difficulty in using
such compilations for quantitative studies is that they contain data
obtained using a variety of heterogeneous techniques.  Thus, not only are
there likely to be systematic errors in the derived masses, but the
nature of these errors will vary within the compilation.

To minimize the impact of such uncertainties, we have chosen to
consider a sample containing only objects from a single study where
the analysis has been performed in a consistent fashion.  Although it
could not be argued that such a sample is necessarily free of
systematic errors, one might reasonably hope that the consistent
analysis should produce relatively consistent results.  For example,
if one galaxy is found to have a more massive black hole than another
optically-identical galaxy within a single sample, then it is likely
that the two systems are intrinsically different; one cannot say the
same if one compares two galaxies from different samples that have
been analyzed using different techniques.

The largest consistently-analyzed sample available is that published
by Magorrian et al.\ (1998).  Their study of galaxies' stellar
kinematics presented estimates for the masses of the central black
holes and spheroidal components of 32 galaxies.  The only exceptional
galaxy in this dataset is NGC~1399.  The luminosity distribution of
this galaxy has a very large diffuse envelope, making it one of the
most extreme known examples of a cD galaxy (Schombert 1986).  It is
therefore almost impossible to disentangle the mass of this extensive
galaxy from the mass of the cluster that surrounds it.  In fact, it is
interesting to note that this galaxy lies well above Magorrian et
al.'s (1998) mean relation between $M_{\rm BH}$ and $M_{\rm sph}$.
However, if one adds to $M_{\rm sph}$ an estimate for the total mass
in the cD envelope (which was excluded from the original mass
estimate), it is straightforward to place NGC~1399 right on the mean
relation.  Unfortunately, it is difficult to justify this ad hoc
correction when one is trying to carry out a consistent analysis.
Given the undesirability of such a posteriori manipulation, and the
fact that such extreme cD systems are likely to have evolved by very
different mechanisms from regular ellipticals, NGC~1399 has been
excluded from the sample, leaving a dataset of 31 galaxies.  We
should, however, note that the presence of this galaxy in the sample
makes no difference to the statistical significance of the conclusions
presented below.

\subsection{Age determinations}

As with the black hole mass estimates, it is important that the galaxy
age estimates be made in as consistent a manner as possible.  We have
therefore used values from the recent catalogue of Terlevich \& Forbes
(2000), which is compiled from a relatively homogeneous dataset of
high-quality absorption line measurements for galaxies (e.g. H$\beta$,
H$\gamma$, [MgFe]).  Using the stellar population model of Worthey
(1994), these line indices can be used to break the age/metallicity
degeneracy, thus giving separate age and metallicity estimates.  

For a few galaxies not in the Terlevich \& Forbes (2000) catalogue,
the same line indices have been measured by Trager et al.\ (1998)
using data of comparable quality.  Combining these measurements with
the Terlevich \& Forbes dataset, one can obtain consistent age
estimates for 23 of the galaxies in the current sample.

The line index measurements come from the galaxies' central regions
and are luminosity weighted.  They are therefore dominated by the last
major burst of star formation.  Thus, the age estimate probably
reflects the time since the galaxy's last major merger event,
which will have induced significant amounts of star formation [see
also Forbes, Ponman \& Brown (1998)].

\begin{figure}
\centering \epsfig{figure=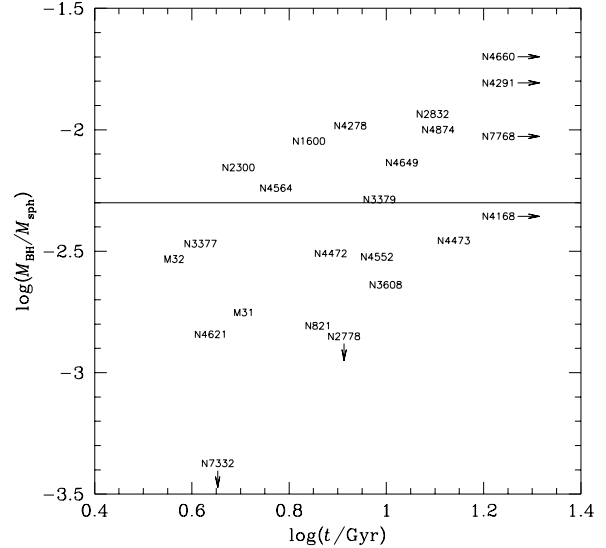, width=8cm}
\caption{ 
Fraction of galaxies' masses in their central black holes as
a function of the ages inferred for their stellar components.
The names of the galaxies are labelled.  The horizontal
line shows the nominal mean value for black hole masses (Magorrian
et al.\ 1998).}
\end{figure}

\subsection{The black hole mass -- galaxy age relation}

Figure~1 shows the fraction of each galaxy's spheroidal component mass
that resides in its central black hole as a function of the age
inferred for the galaxy's stellar population.  There is clearly a
large amount of scatter in this plot; indeed, since there are sizeable
uncertainties in both the black hole mass determinations and the age
estimates, one could not expect to see a tight correlation.  However,
there is a definite trend in the sense that older galaxies of a given
total mass contain more massive black holes: the four youngest
galaxies all have black holes whose masses lie below the mean of
$M_{\rm BH} = 0.005 M_{\rm sph}$, while three of the four oldest
galaxies lie above this line.  More quantitatively, a Spearman rank
test rejects the possibility that $M_{\rm BH}/M_{\rm sph}$ and $t$
are uncorrelated at $>99\%$ confidence.  The robust nature of a rank
test means that the significance of this correlation does not hang on
the outlying points -- the same confidence level is reached if, for
example, NGC~7332 is excluded from the analysis.

\section{Discussion}

Although there does appear to be a significant correlation between
measured black hole mass and galaxy age estimate, it is not
necessarily astrophysical in origin.  We must first consider the
possibility that it arises from some systematic error in the analysis.
However, the kinematic data from which the black hole masses were
inferred are completely independent from the line index data that
provide the age estimates.  Since the line index data were not
selected with this project in mind, and the black hole mass estimates
played no role in the choice of sample, the selection process
cannot have induced the correlation that is seen in Fig.~1.  Further,
the independent nature of the data sets used to measure the two
ordinates means that there can be nothing in this analysis that might
preferentially over-estimate the black hole masses in old galaxies,
or underestimate the masses in young systems.

It should also be borne in mind that the absolute calibrations of the
black hole masses and galaxy ages are significantly uncertain.  In the
case of the absorption line indices, for example, the age estimates
are derived from spectral synthesis modelling, which remains a
somewhat uncertain process, so the absolute values of the ages of two
galaxies may be quite ill-determined.  However, the fact that one is
older than the other can be determined relatively reliably by this
modelling process, so the approximate ordering of galaxy ages can be
determined quite robustly.  Since the Spearman rank test described
above depends only on this ordering, the statistical significance of
the correlation is not dependent on the details of the adopted
calibration.

It would thus appear that there is an underlying astrophysical
correlation between the fraction of a galaxy's mass in its central
black hole and the age of its most recently formed stellar component.
Hence, in addition to the established correlation between black hole
mass, $M_{\rm BH}$ and galaxy mass, $M_{\rm sph}$, there seems to be a
``second parameter'' correlation with the age of the youngest stellar
component.  At any given value of $M_{\rm sph}$, different age
galaxies will have different values of $M_{\rm BH}$, so this secondary
correlation must go some way toward explaining the scatter in the
primary relation.  

We have sought to quantify the contribution of this second parameter
to the scatter in the relation between $M_{\rm sph}$ and $M_{\rm BH}$
by calculating
\begin{equation}
\log(M_{\rm BH}/M_{\rm sph})^* = \log(M_{\rm BH}/M_{\rm sph}) - 
\log(t/10\,{\rm Gyr}).
\end{equation}
This process corrects the mass ratio for the effects of age by
subtracting the simplest possible linear fit to the correlation in
Fig.~1.  As one would expect, this correction reduces the scatter in
the relation: for the data in this sample, the dispersion in
$\log(M_{\rm BH}/M_{\rm sph})$ is 0.42 dex while that in $\log(M_{\rm
BH}/M_{\rm sph})^*$ is only 0.31 dex.  Clearly, even the corrected mass
ratio still contains considerable scatter.  However, given the large
uncertainties in the individual black hole mass and galaxy age
determinations, it would be very surprising if the dispersion were
reduced below a factor of two ($\sim 0.3$ dex).

The simplest explanation for the existence of the second parameter
correlation is that a single physical process couples the growth of
the central black hole to the triggering of star formation in a
galaxy.  As outlined in the Introduction, the hierachical picture of
galaxy and black hole evolution described by Kauffmann \& Haehnelt
(2000) suggests that galaxy mergers lie behind both processes.  Where
the last major merger occurred long ago, it will have taken place in a
gas-rich environment that will provide ample fuel to augment the mass
of the black hole.  Since the last major episode of star formation
will also be triggered in the merger, such galaxies will contain old
stellar populations and massive black holes.  Conversely, galaxies
formed in more recent mergers will contain under-massive black holes
and younger stellar populations.

Although the correlation between black hole mass and galaxy age is
predicted by the hierarchical merging models, it should be borne in
mind that such a correlation is a fairly generic prediction of any
model in which the black hole mass grows over time.  Even if galaxies
form monolithically, those that form first -- and hence contain the
oldest stellar populations -- will have had time to grow the largest
black holes.  The models that do not fit easily with this correlation
are those in which the black holes and stellar components form at
entirely different times -- it would be hard to explain the observed
correlation if, for example, the central black holes were entirely
primordial.

The study of black hole demographics is maturing rapidly, and, as we
hope we have shown, it is already possible to detect phenomena beyond
the basic relation between $M_{\rm BH}$ and $M_{\rm sph}$.  In the
near future, larger sets of both the kinematic and line-strength data
will become available, and more sophisticated modelling techniques
will be developed to refine the estimates for black hole masses and
galaxy ages.  With these tools, it will become possible to address
subtler questions, such as whether a galaxy's environment plays a
significant role in its black hole growth rate.  Such analyses will
provide key tests for theories of black hole formation within the
broader context of galaxy evolution.

\section*{Acknowledgments}

It is a pleasure to thank the referee, Bob Mann, for a range of
helpful suggestions.

\label{lastpage}
\end{document}